\documentclass[aps,prl,superscriptaddress,showpacs]{revtex4}
\usepackage{graphicx}

\begin{document}

\title{Key role of asymmetric interactions in low-dimensional heat transport}
\author{Shunda Chen$^1$, Yong Zhang$^1$, Jiao Wang$^1$, Hong Zhao$^{1,2}$ }
\affiliation{Department of Physics and Institute of Theoretical Physics and Astrophysics,
Xiamen University, Xiamen 361005, Fujian, China}
\affiliation{Collaborative Innovation Center of Chemistry for Energy Materials, Xiamen
University, Xiamen 361005, Fujian, China}
\email{zhaoh@xmu.edu.cn}
\date{\today }
\begin{abstract}
We study the heat current autocorrelation function (HCAF) in
one-dimensional, momentum-conserving lattices. In particular, we explore if
there is any relation between the decaying characteristics of the HCAF and
asymmetric interparticle interactions. The Lennard-Jones model is
intensively investigated in view of its significance to applications. It is
found that in wide ranges of parameters, the HCAF decays faster than
power-law manners, and in some cases it decays even exponentially. Following
the Green-Kubo formula, the fast decay of HCAF implies the convergence of
heat conductivity, which is also corroborated by numerical simulations. In
addition, with a comparison to the Fermi-Pasta-Ulam-$\beta$ model of
symmetric interaction, the HCAF of the Fermi-Pasta-Ulam-$\alpha$-$\beta$
model of asymmetric interaction is also investigated. Our study suggests
that, in certain ranges of parameters, the decaying behavior of the HCAF is
correlated to the asymmetry degree of interaction.
\end{abstract}

\maketitle



\section{Introduction}

\label{intro}

How a fluctuation relaxes in the equilibrium state is very important, not
only in its own right, but also because it governs the transport behavior of
a system in a nonequilibrium state. Relaxation of a fluctuation is
characterized by the corresponding correlation function, and according to
the linear Boltzmann and Enskog equations, if a system is not in a critical
region, in general its correlation functions are believed to decay
exponentially at long times \cite{Hansen, Pomeau}. Consequently, the
integral of a flux autocorrelation function, which appears in the Green-Kubo
formula, converges and thus guarantees a size-independent transport
coefficient~\cite{KuboSP}. Until the late 1960s, the viewpoint that
correlation functions decay exponentially had been prevailing. However,
after Alder and Wainwright~\cite{Alder} numerically evidenced the long-time
tail of velocity autocorrelation in gas models in 1970, extensive analytical
and numerical studies suggested that in one-dimensional (1D) and
two-dimensional (2D) momentum-conserving systems, the heat current
autocorrelation function (HCAF) generally decays in power-law manners
instead. (See Refs.~\cite{Lebowitz, Lepri, Dharrev} for reviews and more
literatures.) The HCAF is defined as 
\begin{equation}
C(t)\equiv\langle J(0)J(t)\rangle,
\end{equation}
where $\langle\cdot\rangle$ denotes the equilibrium thermodynamic average
and $J(t)$ is the total heat current. For 1D momentum-conserving systems, a
recent analytical study has summarized that $C(t)\sim t^{-\gamma}$ with $%
\gamma=1/2$ and $\gamma=2/3$, respectively, for systems with symmetric and
asymmetric interparticle interactions~\cite{Beijeren}. For 2D
momentum-conserving systems~\cite{Lepri, Nara}, the decaying exponent is $%
\gamma=1$. An important consequence of the power-law decay is the divergence
of the heat conductivity for $\gamma \leq 1$ following the Green-Kubo
formula~\cite{KuboSP, Lepri} 
\begin{equation}
\kappa=\frac{1}{k_{B}{T^{2}}d}\lim\limits _{\tau\rightarrow\infty} \lim
\limits _{V\rightarrow\infty}\frac{1}{V}\int_{0}^{\tau}C(t)dt.
\end{equation}
Here $\kappa$, $T$, $d$, and $V$ are, respectively, the heat conductivity,
the temperature, the dimension, and the volume of the system, and $k_B$ is
the Boltzmann constant. It implies that low dimensional (i.e., 1D or 2D)
materials, such as nanowires and graphene flakes, may possess an anomalous
thermal transport property. At present, this viewpoint --- that the HCAF of
low dimensional momentum-conserving systems generally decays in power-law
manners --- is a mainstream viewpoint that has also been accepted by
experimentalists.

Nevertheless, in a recent numerical study~\cite{Zhong12}, it has been found
that in 1D momentum-conserving lattices with asymmetric interparticle
interactions, the heat conductivity may turn out to be independent of the
system size. This finding implies that in such a system, the HCAF may decay
faster than power-law manners, or in a power-law manner but with $\gamma > 1$%
. Later, faster decay has also been observed in several other 1D
momentum-conserving lattices with asymmetric interactions~\cite{Chen12,
Zhong13, Savin}. The faster decay behavior is in clear contrast with
existing theories and has significant importance. On one hand, it requires
us to revisit the theories developed for low dimensional transport problem
during last decades; On the other hand, it implies that realistic low
dimensional materials, such as nanowires and graphene flakes, may still
follow the well-known Fourier heat conduction law~\cite{G14, G15, GMC15},
because real materials usually show the thermal expansion effect, which is a
consequence of asymmetric interparticle interactions. This problem hence
deserves careful studies.

The purpose of this paper is to explore if there exists any link between
asymmetric interparticle interactions and the decaying behavior of the HCAF
in 1D momentum-conserving lattices. The model we focus on is the
Lennard-Jones (L-J) lattices, whose interaction asymmetry degree depends on
both the temperature and a pair of system parameters. We show that in wide
ranges of temperature and system parameters, this model has a high asymmetry
degree and the HCAF decays faster than any power-law manners. Nonequilibrium
simulations also confirm that it has a size-independent heat conductivity.
The Fermi-Pasta-Ulam-$\alpha$-$\beta$ (FPU-$\alpha$-$\beta$) model is also
studied. This model has asymmetric interaction but faster decay of its HCAF
has not been reported yet~\cite{Savin, Das, Dhar, WL}. Our analysis shows
that in clear contrast to intuition, the effects of asymmetric interactions,
such as thermal expansion, do not vanish in the low temperature regime;
rather, it becomes even stronger in some sense. In this regime, fast decay
of the HCAF, that may lead to normal heat conduction, is observed as well.
To make a comparison, we also study the HCAF of the Fermi-Pasta-Ulam-$\beta$
(FPU-$\beta$) model of symmetric interaction, in such a low temperature
regime. Our results suggest that under proper conditions, asymmetric
interactions may lead to the fast decay of the HCAF, i.e., certain
correlation may exist between fast decay of the HCAF and asymmetric
interactions. This finding could be a helpful clue for further studies on
the HCAF. As the two lattice models of asymmetric interactions (L-J model
and FPU-$\alpha$-$\beta$) are general in some sence, we suspect such
correlation may also be found in other asymmetric lattices with proper
parameter values. Nevertheless, we point out that at present the exact
conditions for observing such correlation are not clear yet, and it would be
risky to interpret such correlation as a cause-effect relationship. These
problems deserve more efforts and should be clarified in future.

We adopt molecular dynamics studies for our aim here. At present numerical
analysis plays an important role in studying the heat conduction properties
of low dimensional systems. Although the problem can be dealt with
analytically, certain approximations and assumptions have to be resorted to.
For example, in analytical studies, it has been generally assumed that all
slow variables of relevance for the long-time behavior of hydrodynamics and
the related time correlation functions are the long-wavelength Fourier
components of the conserved quantities' densities~\cite{Beijeren}. For this
reason, particular attention must be paid when analytical results are
compared with simulation and experiment results~\cite{Das}. As to
experimental studies, despite the fact that in recent years it has become
technically feasible to measure the heat conductivity of low dimensional
materials, the accessible precision is still far away for drawing convincing
conclusions. In addition, realistic materials studied in laboratories, such
as nanowires and graphene flakes, may not be genuine 1D and 2D objects,
considering their possible transverse motions. Hence, how to control or
evaluate the effects of the transverse motions turns out to be a new
experimental challenge. In contrast, the molecular dynamics method does not
suffer from any of these problems. However, it has its own difficulty: the
finite-size effect, which often makes it hard to reach consensus on the
simulation results~\cite{Chen14}. Inevitably, our study in this paper will
also face the doubt whether our results can be extrapolated to the
thermodynamic limit. Regarding to this concern, by taking the FPU-$\alpha$-$%
\beta$ model as an example, we will show that, even though the Fourier heat
conduction behavior observed with finite system sizes does not hold up to an
infinite system eventually, it does hold up to a physically meaningful
macroscopic system.

In the following we shall first describe the lattice model of Lennard-Jones
potential and present the simulation results; then we shall turn to the
asymmetric, Fermi-Pasta-Ulam-$\alpha$-$\beta$ (FPU-$\alpha$-$\beta$) model.
A brief summary and discussion will be presented in the last section.

\section{Lennard-Jones Lattices}

The Lennard-Jones (L-J) potential has been widely adopted in modeling
realistic materials. It is asymmetric with respect to the equilibrium point,
and our study has shown that in 1D lattices with L-J potentials, the HCAF
can decay faster than power-law manners~\cite{Chen12}. This finding is in
clear contrast to the well known theoretical prediction of the power-law
decay.

The Hamiltonian of a 1D lattice with the nearest neighboring coupling can be
written as 
\begin{equation}
H=\sum_{i}\frac{p_{i}^{2}}{2m_{i}}+U(x_{i}-x_{i-1}-1),
\end{equation}
where $p_{i}$, $x_{i}$, $m_{i}$, and $U$ represent, respectively, the
momentum, the position, the mass of the $i$th particle and the potential
between two neighboring particles. For both models we study in the
following, we assume that all the component particles are identical and have
unit mass; i.e., $m_{i}=1$. The lattice constant is set to be unity so that
the system length $L$ equals the particle number $N$. In our simulations for
the HCAF, the periodic boundary condition is imposed, and the total momentum
of the system is set to be zero. Note that in 1D lattices, if there is no
steady motion (i.e., the total momentum is zero), then the heat current
equals the energy current~\cite{Lepri}; Hence our results can be extended to
the energy current straightforwardly. We consider the total heat current
defined as $J\equiv\sum_{i}\dot{x}_{i}\frac{\partial U} {\partial x_{i}}$ 
\cite{Mai}. To numerically measure the heat current in the equilibrium
state, the system is first evolved from an appropriately assigned random
initial condition for a long enough time ($>10^{8}$ in our simulations) to
ensure that it has relaxed to the equilibrium state, then the total heat
current at ensuing times is measured. The energy density, or the energy per
particle, denoted by $\langle E_i\rangle$, is determined by the initial
condition and is conserved during the simulation. At the low temperature
regime, by using the Viral theorem, we have $k_B T \approx \langle E_i
\rangle \approx 2\langle U_i \rangle$, where $\langle U_i\rangle$ is the
averaged potential energy per particle at the equilibrium state. (Note that
this relation holds only when the harmonic term dominates the potential;
hence only applies at the low temperature regime.) The Boltzmann constant is
set to be unity throughout.

\begin{figure*}[tbp]
\centering
\center\includegraphics[scale=0.92]{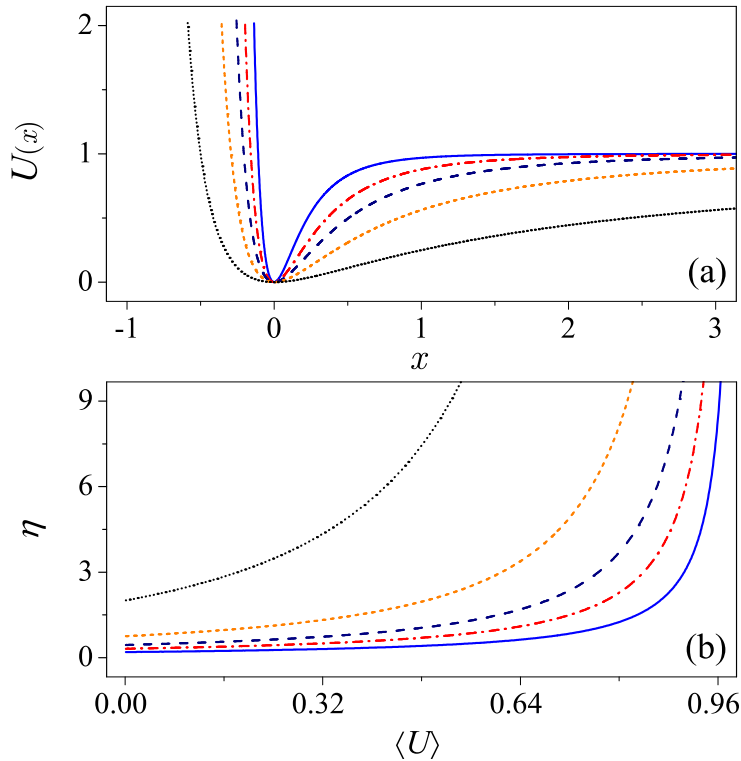}
\caption{Plots of the Lennard-Jones potential function (a) and its asymmetry
degree (b) at various parameter sets. From top to bottom in (a) but
reversely in (b), the curves are for $(m,n)$=(12,6), (8,4), (6,3), (4,2),
and (2,1), respectively.}
\label{fig1}
\end{figure*}

The L-J potential we consider has the form $U(x)=[({x+1})^{-m}-2({x+1}%
)^{-n}+1]$. It involves a pair of parameters, $m$ and $n$, that control the
asymmetry degree of the potential. Without loss of generality, in the
following we fix $m=2n$ so that the minimum of $U(x)$ is fixed at $x=0$. The
potential is asymmetric with respect to the equilibrium point [see Fig.
1(a)]. In order to compare the asymmetry degree for different $(m,n)$ and at
different temperatures, we introduce the following measure of the asymmetry
degree: 
\begin{equation}
\eta \equiv {\frac{d }{{d\left\langle U \right\rangle }}}({x_+ } - |{x_- }|);
\end{equation}
where $x_+$ and $x_-$ are, respectively, the right- and the left-side zero
point of $U(x)-\langle U \rangle=0$, with $\langle U \rangle $ being the
average potential energy between two neighboring particles, or the average
potential energy per particle. Note that ${x_+ }-|{x_- }|$ represents
expansion and $\eta$ is equivalent to the thermal expansion coefficient upon
a factor of the heat capacity~\cite{Kittel}. As $\eta$ captures and reflects
thermal effects of asymmetric interactions, it is a natural and physically
meaningful choice to measure the asymmetry degree. As Fig. 1(b) shows, the
asymmetry degree of the L-J potential increases as the average potential
energy $\langle U \rangle$, and for a fixed $\langle U \rangle$ value it
increases as the parameter pair $(m,n)$ varies from $(m,n) = (12,6)$ to $%
(m,n) = (2,1)$.

\begin{figure*}[tbp]
\centering
\center\includegraphics[scale=1.2]{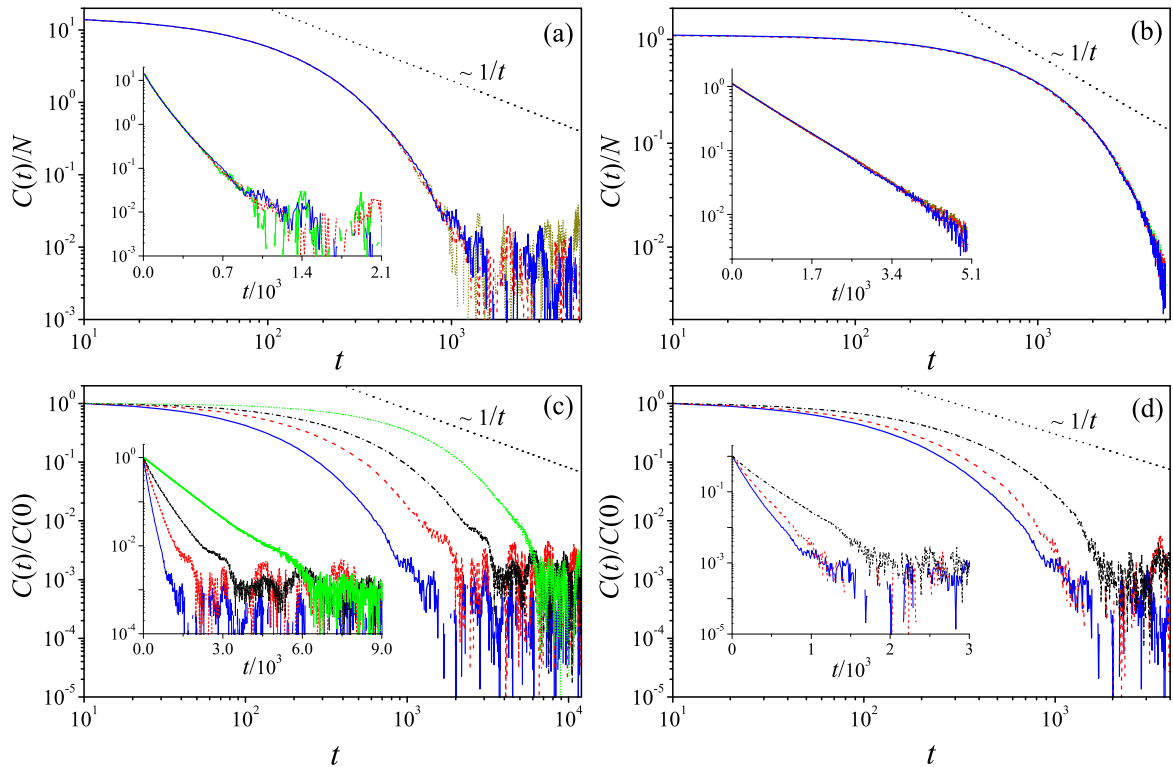}

\caption{The autocorrelation function of the heat flux in the 1D
Lennard-Jones lattices. The inset in each panel is the same as the main
panel but in the semi-log scale as a comparison. In (a), the results for $%
(m,n)$=(12,6) and $\langle E\rangle=0.5$ are presented with $N=4096$ (short
dotted line), 8192 (dashed line), and 16384 (solid line), respectively. The
black dotted line indicates the scaling $\sim1/t$ for reference. (b) The
same as (a) but for $(m,n)$=(2,1) with $N=16384$ (short dotted line), 32768
(dashed line), and 65536 (solid line). (c) A comparison between the results
for $(m,n)$=(12,6), (8,4), (6,3), and (2,1) (from bottom to top) with $%
N=16384$ and $\langle E\rangle=0.5$. (d) A comparison between the results
for various energy density $\langle E\rangle=0.5$, 1, and 2 (from bottom to
top), with $(m,n)$=(12,6) and $N=16384$.}
\label{fig2}
\end{figure*}

We will focus on the case of $(m,n) = (12,6)$, the most frequently adopted
parameters in literatures, but also discuss other values of $(m,n)$ when it
is in order. In Fig. 2 we show the simulation results of the HCAF. We have
performed the finite-size effect analysis as outlined in Ref.~\cite{Chen14}
and found that in 1D L-J lattices, the finite-size effect is in fact
negligible, which is very favorable for numerical studies of the HCAF.
Thanks to this property, the asymptotic decaying behavior of $C(t)$ can be
reliably revealed even with a comparatively small system of $N \approx 4
\times {10^3}$ ( $N$ is the total number of particles in the system). To
show this property, in Fig. 2(a) and (b) the HCAF at different system sizes
are compared for $(m,n) = (12,6)$ and $(m,n) = (2,1)$, respectively. The
energy density $\langle E\rangle $, i.e., the average energy per particle,
is fixed at $\langle E\rangle=0.5$, which corresponds to the temperature of $%
T \approx 0.55$ and the average potential energy per particle $\langle
U\rangle \approx 0.2$. It can be seen that in both cases, all the curves of $%
C(t)$ perfectly collapse onto one upon scaling by the system size $N$. This
evidence strongly suggests that the finite-size effects are negligible for $%
N > 4 \times {10^3}$. Note that the oscillating tails around zero for $t > {%
10^3}$ appear in Fig. 2(a) [also in Fig. 2(c) and Fig. 2(d)] are due to
uncertainty of statistical average of $J(0)J(t)$ [see Eq. (1)], and the
negative parts of the oscillations are not shown as the vertical axis is in
the logarithmic scale.

It shows clearly in Fig. 2(a) that the HCAF decays faster than any power-law
manners. The inset presents the data in the semi-log scale, from which one
can see that the decay manner is already very close to an exponential one.
Fig. 2(b) shows the results for $(m,n) = (2,1)$, and $C(t)$ curves can be
regarded as decaying exponentially with certainty. It can be found from Fig.
1(b) that the L-J potential with $(m,n) = (2,1)$ has relatively higher
asymmetry degree; we thus conjecture the higher the asymmetry degree is, the
closer the decaying behavior of the HCAF tends to be exponential. To check
this conjecture, in Fig. 2(c) we compare the HCAF for various $(m,n)$
values. It shows that, as $(m,n)$ varies from $(12,6)$ to $(2,1)$, i.e., as
the asymmetry degree increases [see Fig. 1(b)], the $C(t)$ curve in the
semi-log scale becomes straighter and straighter [see the inset of Fig.
2(c)], suggesting that the decaying behavior indeed tends to be exponential
progressively.

As Fig. 1(b) shows, for a given parameter pair $(m,n)$, the asymmetry degree
is controlled by the average potential energy per particle $\langle U\rangle$%
: it increases dramatically as the latter. As the temperature monotonically
increases with $\langle U\rangle$, it implies that the temperature could
also be correlated to the decaying behavior of the HCAF: For the L-J model
at a higher temperature, the average potential per particle is higher and
consequently, the potential's asymmetry degree would be higher [see Fig.
1(b)]. In Fig. 2(d) the temperature dependence of the decaying behavior of
the HCAF is studied for $(m,n) = (12,6)$ with $\langle E\rangle $=0.5, 1,
and 2. For these $\langle E\rangle $ values, the corresponding temperature
is $T \approx $0.55, 1.2, and 2.4, and the corresponding $\langle U\rangle $
is about 0.2, 0.4, and 0.8. Again, the $C(t)$ curve in the semi-log scale
[see the inset of Fig. 2(d)] becomes straighter and straighter and at $%
\langle E\rangle $=1, it has already become a perfect exponential function.
This again confirms that the decaying behavior of the HCAF is correlated to
the asymmetry degree. Following the Green-Kubo formula [see Eq. (2)], the
fact that $C(t)$ decays faster than the power law of $\sim{t^{-1}}$ suggests
that the integral of $C(t)$ is convergent. In addition, for $N > 4 \times {%
10^3}$, the fact that $C(t)/N$ curves for different system sizes agree with
each other as shown in Fig. 2(a) and (b) implies that the thermal
conductivity does not depend on the system size. In Fig. 3, we present the
heat conductivity given by the Green-Kubo formula. Note that in order to get
rid of the possible finite-size effect, a practical procedure for obtaining
the heat conductivity at a given system size $N$, denoted by ${\kappa _{GK}}%
(N)$, is to truncate the time integration in the Green-Kubo formula~\cite%
{Lepri, Pros} up to the time ${\tau _{tr}}(N) = N/(2{v_s})$ (${v_s}$ is the
sound speed of the system)~\cite{Chen14}; i.e., 
\begin{equation}
\kappa_{GK}(N)=\frac{1}{k_{B}{T^{2}}}\int_{0}^{\tau}\frac{C(t)}{N}dt.
\end{equation}
It can be seen from Fig. 3 that $\kappa_{GK}$ increases as $N$ for $%
N<4\times 10^3$ but gets saturated for larger $N$, which is in good
consistence with the results of $C(t)/N$ [see Fig. 2(a)].

\begin{figure*}[tbp]
\centering
\center\includegraphics[scale=1.5]{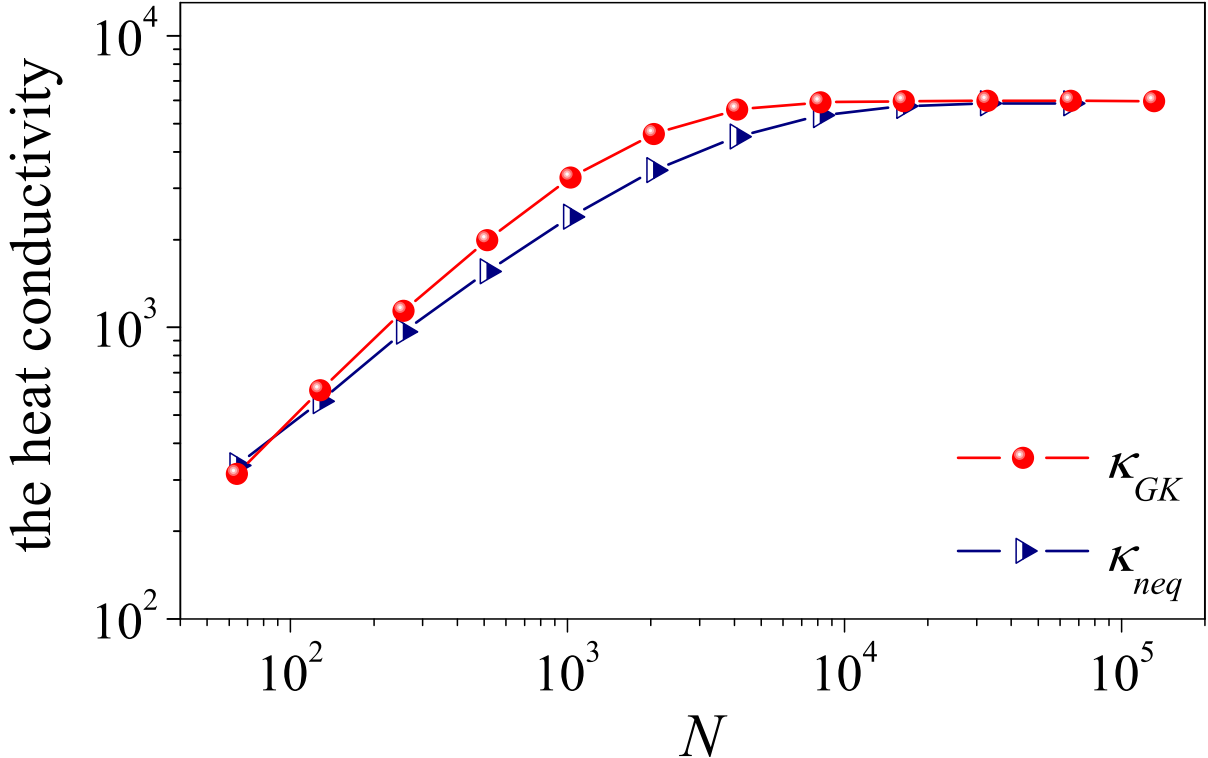}
\caption{The system size dependence of the heat conductivity obtained by
using the Green-Kubo formula ($\protect\kappa_{GK}$) and the nonequilibrium
settings ($\protect\kappa_{neq}$) for the Lennard-Jones lattices with $(m,n)$%
=(12,6). In the simulations of $\protect\kappa_{GK}$, the average energy per
particle is set to be $\langle E\rangle=0.5$, corresponding to a system
temperature of $T\simeq0.55$. In the simulations of $\protect\kappa_{neq}$,
the temperatures of the heat baths are set to be $T_{+}=0.7$ and $T_{-}=0.4$
so that the average temperature of the system is 0.55 as well.}
\label{fig3}
\end{figure*}

To further verify the convergence of the heat conductivity, we have also
computed it directly by using the nonequilibrium simulations: Two Langevin
heat bathes~\cite{Dharrev} with with temperatures $T_{+}=0.7$ and $T_{-}=0.4$%
, respectively, are connected to the two ends of an L-J lattice. After the
stationary state is established, the heat conductivity is evaluated by
assuming the Fourier law $j=-\kappa dT/dx$, where $j\equiv \langle J\rangle/N
$, and $dT/dx \equiv ({T_+ } - {T_- })/N$ is the temperature gradient along
the lattice. The results (denoted by $\kappa_{neq}$) are also presented in
Fig. 3 for a close comparison. Again, the heat conductivity measured in this
way tends to saturate and the value it tends to is the same as that obtained
by integrating the HCAF with the Green-Kubo formula.

To summarize, 1D L-J lattices obey the Fourier law at large system sizes,
and their HCAF decay faster than power-law manners. This conclusion is not
affected by the finite-size effects.

\begin{figure*}[tbp]
\centering
\center\includegraphics[scale=1]{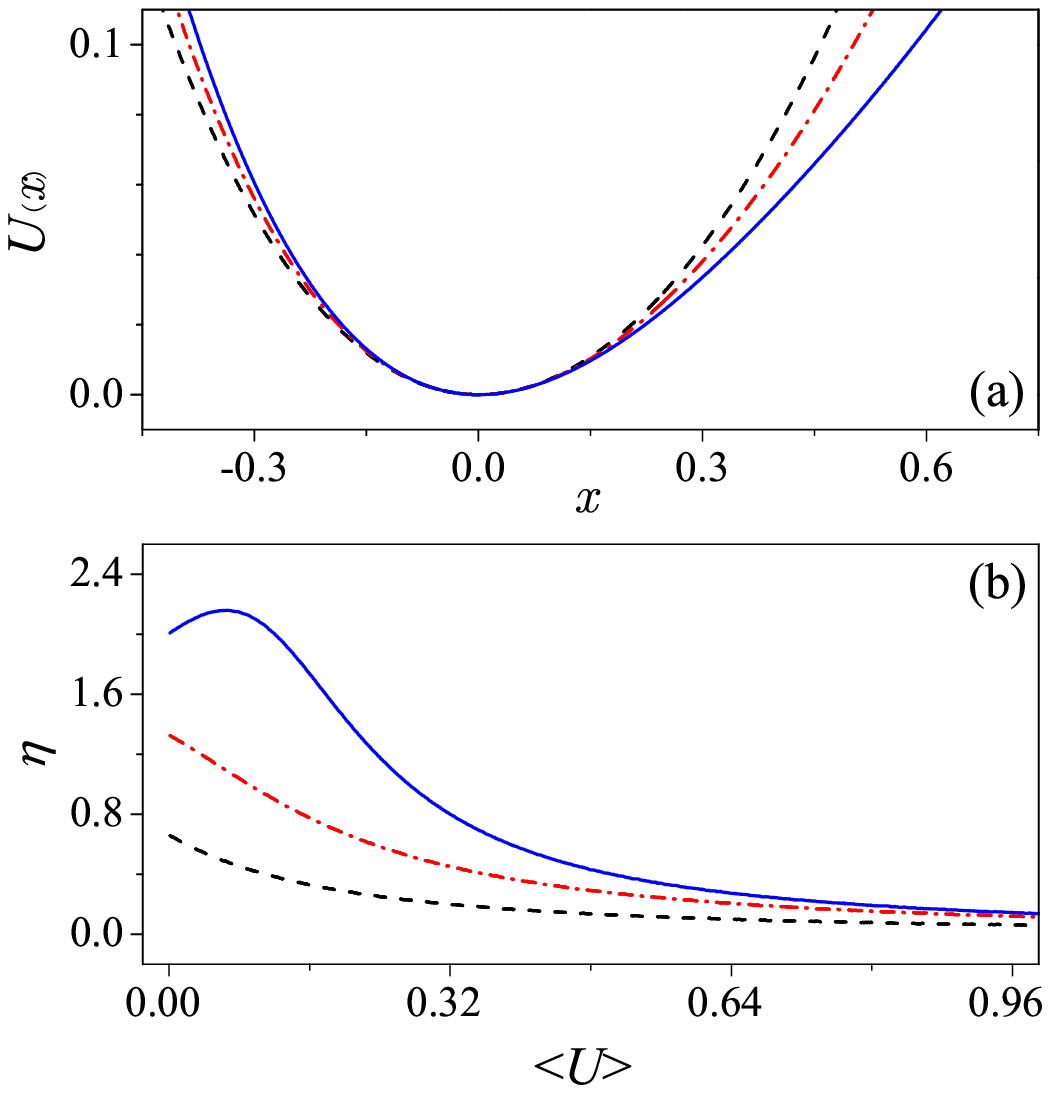}
\caption{Plots of the potential function of the FPU-$\protect\alpha$-$%
\protect\beta$ model (a) and its asymmetry degree (b) at various parameters.
In both panels, the dashed, the dot-dashed, and the solid curve is for $%
\protect\alpha=0.5$, 1, and 1.5, respectively; $\protect\beta=1$ in all the
cases. }
\label{fig4}
\end{figure*}

\section{FPU-$\protect\alpha$-$\protect\beta$ lattices}

The potential of the 1D FPU-$\alpha$-$\beta$ model is $U(x) = {x^2}/2 -
\alpha {x^3}/3 + \beta {x^4}/4,$ where the two parameters $\alpha$ and $\beta
$ determine, respectively, the asymmetric and the symmetric nonlinear term.
As our aim is to investigate the effects induced by the asymmetric term, we
fix $\beta=1$ throughout. Fig. 4(a) shows the potential for three different
values of $\alpha$, and Fig. 4(b) shows the corresponding result of the
asymmetry degree.

From Fig. 4(b) it can be seen that the FPU-$\alpha$-$\beta$ model is quite
different from the L-J model: Though in general the asymmetry degree
increases as $\alpha$, it generally decreases and tends to zero as the
average potential energy $\langle U\rangle $ increases. In addition, for
larger $\alpha$ the asymmetry degree may depend on $\langle U\rangle $ in a
nonmonotonic manner as in the case of $\alpha$=1.5. At the first glance,
these results seem to contradict to our intuition; e.g., when the
temperature tends to zero, $\langle U\rangle $ decreases accordingly, and
the potential energy represented by the quadratic term in the potential
function would become dominant. For this reason, one may expect that the
physical properties of the FPU-$\alpha$-$\beta$ model would converge to
those of a harmonic lattice. However, we would like to point out that, for
some nonlinear effects, such as thermal expansion, the cubic term in the
potential function is always the leading term; hence whether a nonlinear
effect is non-negligible in the low temperature limit depends on if it could
stand out from the linear effect. As far as thermal expansion is concerned,
in the low temperature limit the thermal expansion coefficient tends to be a
temperature independent constant proportional to $g/{c^2}$ with $g$ and $c$
being, respectively, the coefficient of the cubic and the quadratic term of
the Taylor expansion of the given potential function (see Ref.~\cite{Kittel}%
). For the FPU-$\alpha$-$\beta$ model $g/{c^2} = 4\alpha /3$; i.e., the
larger is $\alpha$, the larger is the thermal expansion coefficient, which
is in good consistence with the results of the asymmetry degree given in
Fig. 4(b). [Note that for the L-J model, the thermal expansion coefficient
does not tend to zero in the low temperature limit either, and the value of $%
g/{c^2}$ for various parameter $(m,n)$ is also in consistence with the
results given in Fig. 1(b).]

If the decaying behavior of the HCAF is correlated to the asymmetry degree,
then based on Fig. 4(b), we could expect that faster decay may be observed
in the low temperature regime where the asymmetry degree is high enough. In
the high temperature regime, the asymmetry degree tends to be low, which is
consistent with the fact that as the temperature increases, the symmetric
quartic term becomes overwhelmingly dominating so that the expansion
quantity ${x_+ } - |{x_- }|$ tends to be a constant, inducing in turn a
decreasing thermal expansion rate. As a result, one may expect instead slow
decay of the HCAF.

Fig. 4(b) also tells that the asymmetry degree is relatively strong only at
low temperatures, in particular for the potential with a smaller value of $%
\alpha$. To reveal the effects of the asymmetric potential, we first
investigate the HCAF at a very low energy density of $\langle E\rangle = 0.05
$. For this energy density, we have $T \approx 0.05$ and $\langle U\rangle
\approx 0.025$. In Fig. 5(a) we show the HCAF for $\alpha$=0.5, 1, and 1.5
with $N = 8192$; it can be seen that for $t < 2 \times {10^4}$, before the
fluctuations (due to uncertainty of statistical average) set in, $C(t)$
decays more rapidly --- in an exponential-like way --- than any power-law
manners in all three cases. In addition, we have also verified that for all
these cases $C(t)/N$ does not change as $N$ as long as $N > {10^3}$.

\begin{figure*}[tbp]
\centering
\center\includegraphics[scale=1]{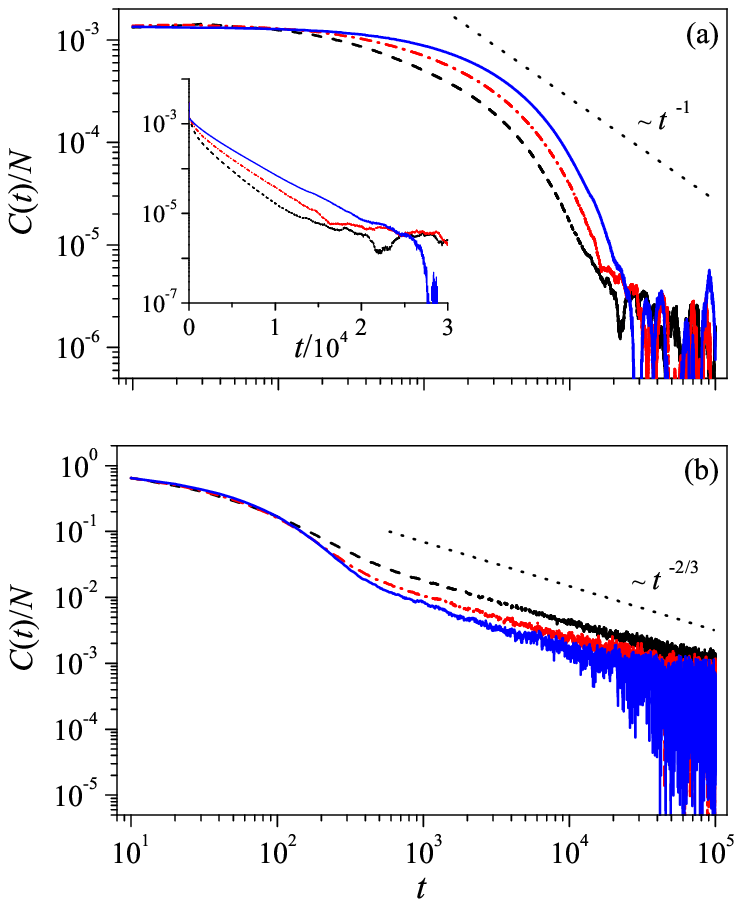}

\caption{The autocorrelation function of the heat flux in the FPU-$\protect%
\alpha$-$\protect\beta$ model for $N=8192$ and $\protect\beta=1$. In both
panels, the dashed, the dot-dashed, and the solid curve is for $\protect%
\alpha=0.5$, $1$, and $1.5$, respectively. Panel (a) is for the low energy
density (temperature) case with $\langle E\rangle=0.05$, $T\simeq0.05$ and $%
\langle U\rangle\simeq0.025$; The inset is the same but in the semi-log
scale. Panel (b) is for the high energy density (temperature) case with $%
\langle E\rangle=0.8$, $T\simeq0.9$ and $\langle U\rangle\simeq0.3$.}
\label{fig5}
\end{figure*}

Now let us turn to the case of a high temperature: $\langle E\rangle=0.8$,
which corresponds to $T \approx 0.9$ and $\langle U\rangle \approx 0.3$.
Compared with the low temperature case, it can be seen from Fig. 4 (b) that
the asymmetry degree has greatly diminished. In this case, the HCAF does
show a power-law tail with the decaying exponent close to 2/3 [see Fig.
5(b)] as predicted by the self-consistent mode coupling theory~\cite{Del06,
Del07}.

Fig. 5(b) also suggests that the HCAF undergoes an initial faster decaying
stage which lasts longer and longer with the increase of $\alpha$. This is a
signal that the asymmetric potential term even plays a role in the high
temperature regime; i.e., higher asymmetry degree can maintain a longer
initial faster decaying stage. It can be expected that if the temperature is
increased further, the effects induced by the asymmetric potential will
become even more unnoticeable due to the rapid decrease of the asymmetric
degree. This has been verified by our simulations.

Concerning the results given in Fig. 5(a), one may wonder if there exists a
crossover from the exponential-like decay to the power-law decay even at low
temperatures, but the time corresponding to the crossover point, denoted by $%
t^*$, is so long that has gone beyond the scope of our simulations. To study
this question, we computed the HCAF at several low temperatures with $%
\alpha=1$ and $N=8192$ [see Fig. 6(a)]. It can be seen that such a crossover
may exist and $t^*$ may increase very fast as the temperature decreases.

Nevertheless, we would like to argue that, even though the HCAF turns out to
have a power-law tail, \textit{in practice} the faster decay before the
power-law tail can still guarantee an effective constant heat conductivity~%
\cite{Chen13}, as long as the faster decaying stage lasts sufficiently long.
To show this, we calculate the heat conductivity for a finite system size $N$
by using the Green-Kubo formula and divide the integral of $C(t)$ into two
parts; i.e., 
\begin{equation}
\kappa = {\frac{1 }{{{k_B}{T^2}}}}[\int_0^{{\tau _e}} {{\frac{1 }{N}}}
C(t)dt + \int_{{\tau _e}}^{{\tau _{tr}}(N)} {{\frac{1 }{N}}} C(t)dt].
\end{equation}
In the time range of the second integral, i.e., $({\tau _e},{\tau _{tr}}(N))$%
, the HCAF is assumed to decay as $C(t)\sim{t^{ - 2/3}}$. Taking the case of 
$\alpha = 1.5$ as shown in Fig. 5(a) as an example, where one can see that
the initial rapid decay lasts up to ${\tau _e} \approx 2.5 \times {10^4}$
and $C({\tau _e})/N$ drops to about ${10^{ - 6}}$; then the first integral $%
\int_0^{{\tau _e}} {(C(} t)/N)dt \approx 4$ and the second integral $\int_{{%
\tau _e}}^{{\tau _{tr}}} {(C} (t)/N)dt \approx 2.6 \times {10^{ - 3}}{t^{1/3}%
}|_{{\tau _e}}^{{\tau _{tr}}}$. It follows that the contribution of the
power-law tail (given by the second integral) to the heat conductivity is
not comparable to that of the faster decaying part until the system size
reaches up to $N = {10^9}$, given that~\cite{Lepri, Chen14, Pros} ${\tau
_{tr}}(N) \approx N$. In other words, if we suppose that the average
distance between two neighboring particles is one angstrom, then the heat
conductivity would keep in effect unchanged over a wide system size range
from about one micrometer to ten centimeters. So as long as the faster decay
stage $(0,{\tau _e})$ lasts long enough, as evidenced in Fig. 5(a), even
though we assume that the HCAF has a power law tail, its influence to the
heat conduction can still be safely neglected.

\begin{figure*}[tbp]

\centering
\center\includegraphics[scale=1]{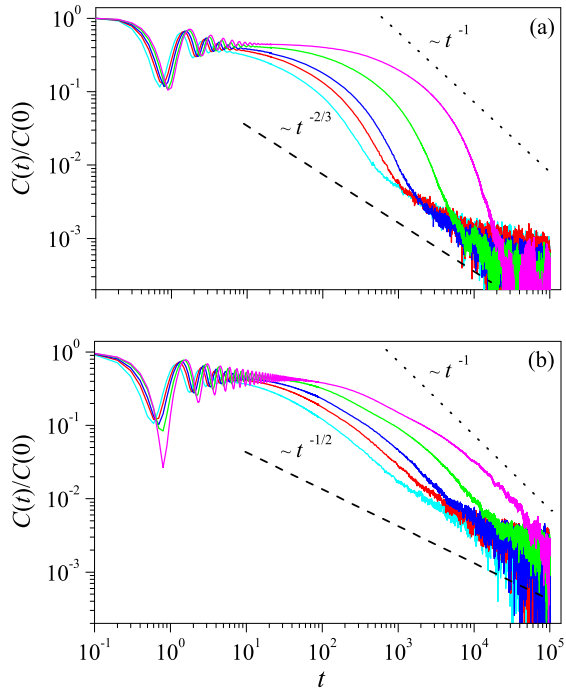}
\caption{The autocorrelation function of the heat flux in the FPU-$\protect%
\alpha$-$\protect\beta$ model with $\protect\alpha=0.5$ (a) and in the FPU-$%
\protect\beta$ model (b). In both panels, solid lines from bottom to top
(see the center part) are for $\langle E\rangle=0.5$, 0.3, 0.2, 0.1 and
0.05, respectively, corresponding to the temperature of $T\simeq0.56$, 0.32,
0.21, 0.1, and 0.05 for both models. For all the cases $\protect\beta=1$ and 
$N=8192$. The dotted lines indicate the scaling threshold $\sim t^{-1}$ for
a convergent Green-Kubo integral and the dashed lines indicate the scaling
of the theoretically predicted power-law tail for one-dimensional,
momentum-conserving systems of asymmetric ($\sim t^{-2/3}$) and symmetric ($%
\sim t^{-1/2}$) interactions. }
\label{fig6}
\end{figure*}

Above analysis raises a relevant question: If these properties remain in the
FPU-$\beta$ model. In Fig. 6(b), we show the HCAF at several temperatures
for the FPU-$\beta$ model with $N=8192$. Roughly, its behavior looks similar
as in the FPU-$\alpha$-$\beta$ model [see Fig. 6(a)], in that there is a
fast decay stage followed by a slow power-law tail. However, qualitative
difference exists: for FPU-$\beta$ model, the decay behavior of HCAF is of
power-law rather than exponential-like, and is slower than $C(t)\sim{t^{ - 1}%
}$ throughout [see Fig. 6(b)]. As a result, the thermal conductivity for FPU-%
$\beta$ model would increase with the system size significantly.

\section{Summary and Discussion}

In summary, we have numerically studied the HCAF in the 1D L-J lattices. It
has been shown that the HCAF generally decays faster than power-law manners.
In addition, we have observed a correlation between the decaying behavior of
the HCAF and the interaction asymmetry degree: As the system parameters $%
(m,n)$ change from $(m,n)=(12,6)$ to (2,1), the asymmetry degree increases,
and meanwhile the HCAF tends to decay more and more exponentially. In
particular, for $(m,n)=(2,1)$, the HCAF shows clear signal of exponential
decay. On the other hand, the asymmetry degree increases as the average
potential energy per particle, or equivalently the temperature. Again, the
HCAF tends to decay exponentially as the temperature increases.

We have also studied the decaying behavior of the HCAF in the FPU-$\alpha$-$%
\beta$ model. While the asymmetry degree increases with the asymmetry
parameter $\alpha$, its dependence on temperature is quite different from
that of the L-J model: in the high temperature regime, the asymmetry degree
decreases monotonically as the average potential energy per particle
increases, in consistence with the fact that the symmetric quartic term in
the potential becomes dominating. We show that in the low temperature regime
the exponential-like fast decay of the HCAF can last for a significantly
long time though a power-law tail may follow.

One important question that cannot be definitely answered via numerical
simulations only is whether the power-law tail would show up if the system
size is out of the scope accessible to simulations. Our observation is that
there are not any signals of the power-law tail in the L-J model with all
the system parameters investigated. However, as to the FPU-$\alpha$-$\beta$
model, we cannot exclude such a possibility based on our present simulation
results.

Nevertheless, we have shown that for the FPU-$\alpha$-$\beta$ model at a low
temperature, even if the HCAF has a power-law tail eventually, effective
constant heat conductivity can still be expected if only the initial faster
decay lasts long enough. In particular, if the HCAF keeps decaying faster
over more than three orders, the contribution of the assumed power-law tail
to the heat conductivity can be neglected even when the system reaches a
macroscopic scale. Namely, the slow tail of the HCAF may not necessarily
imply abnormal transport \textit{in practice}, hence should be carefully
analyzed when theoretical predictions based on the power-law tail are
applied to experiments. (It is worth noting that though for the FPU-$\beta$
model there is also a fast decaying period in low-temperature regime, the
decaying rate is slower than $\sim1/t$.) Based on these analysis, we
conclude that with proper system parameters, both the L-J model and the FPU-$%
\alpha$-$\beta$ model (at a sufficient low temperature) may have the normal
heat conduction property given by the Fourier law. As realistic materials
usually contain asymmetric interactions, we think this result may be
significant to applications as well.

In a recent work, heat conduction of 1D L-J lattices was studied~\cite{Savin}
at different particle densities. Though the authors did not show how the
HCAF decays, they provided the convergent heat conductivity at low
temperature regime by using the Green-Kubo formula, which implies that the
HCAF decays faster than $\sim1/t$. This is consistent with our earlier
conclusion on this model~\cite{Chen12}. In that work~\cite{Savin}, the
authors also conjectured that the lattices with the parabolic and/or quartic
confining potential, to which the Fermi-Pasta-Ulam model belongs, should all
exhibit anomalous heat transport. Our study suggests that their conjecture
should be studied further because as we have shown here, the heat conduction
behavior of the FPU-$\alpha$-$\beta$ model can also be normal. In addition,
normal heat conduction has also been evidenced in other lattices with
confining potential~\cite{Zhong12, Chen12, Zhong13}, such as the piecewise
parabolic potential~\cite{Zhong13}. Therefore, more efforts are still needed
to unveil the exact conditions under which the HCAF decays faster.

Finally, we emphasize that asymmetric interactions are not a sufficient
condition for the faster decay of the HCAF and normal heat conduction, just
as suggested by the results of the FPU-$\alpha$-$\beta$ model at high
temperatures. Asymmetric interactions may not be a necessary condition for
the faster decay of the HCAF and normal heat conduction, either. To this
end, the 1D rotator lattice ~\cite{Gia00, GS00} is the only known example.
This model has normal heat conductivity under certain conditions, but its
interaction is symmetric. However, recent studies suggest that the angle
variables of this model do not constitute a conserved field \cite{HSpohn14,
DD14}, implying that this model is in effect irrelevant with the subject we
discuss here. It is worth noting that if this example is excluded, then all
the 1D momentum-conserving systems where the HCAF has been reported to decay
faster up to now have asymmetric interactions~\cite{Zhong12, Chen12,
Zhong13, Savin}. Recently, two more model systems have also been added to
this category: 1D hard-point gas and 1D Toda lattice with alternative masses~%
\cite{Fourier14}. The faster decay of the HCAF and normal heat conduction
are found in both models in certain range of the system size when the mass
ratio tends to unity, i.e., when the systems approach their integerable
limit. Taking into account all these studies, the faster decay of the HCAF
seems to be a collective consequence of some different factors to which
asymmetric interactions belong.

\vspace{30pt} \noindent\textbf{Acknowledgments} \vspace{20pt}

\noindent Very useful discussions with S. Lepri, R. Livi and A. Politi are
gratefully acknowledged. This work is supported by the NSFC (Grants No.
11335006, No. 11275159, No. 10805036, and No. 11535011).

\end{document}